# Crisis-induced differences in attention towards Ukraine in Twitter 2008-2023

Mark Mets, [1, 2, *] Peter Sheridan Dodds, [4, 5, 6, 7] and Maximilian Schich [3, *]

[1] *School of Humanities, Tallinn University, Tallinn, Estonia*
[2] *Estonian Literary Museum, Tartu, Estonia*
[3] *Baltic Film, Media and Arts School, Tallinn University, Tallinn, Estonia*
[4] *Computational Story Lab, Vermont Complex Systems Institute, MassMutual Center of Excellence in Complex Systems and Data Science, University of Vermont, Burlington, USA*
[5] *Department of Computer Science, University of Vermont, Burlington, USA*
[6] *Santa Fe Institute, Santa Fe, USA*
[7] *Complexity Science Hub, Vienna, Austria*
* *Corresponding authors: M.M. (mark.mets@tlu.ee) & M.S. (mxs@tlu.ee)*

## Abstract

Aggression against Ukraine has drawn widespread international attention, particularly in the wake of the two Russian invasions into Ukrainian territory in 2014 and 2022. Although previous studies have examined social-media dynamics around these events, a comparative longitudinal data-driven view across languages is still missing. This article fills this gap by mapping added attention to "Ukraine" on Twitter in 28 languages from 2008 to 2023, using a deceptively simple DNA microarray–inspired cartography of log over-expression relative to each language's baseline frequency. This macro-scale visualization makes familiar events stand out while uncovering subtler patterns beyond the cognitive reach of any single-language audience. Most strikingly, two nearly non-overlapping language clusters emerge, one peaking around 2014 and the other around 2022 with distinct onset and decay profiles that are consistent with markedly different geopolitical attention regimes around the two invasions. By capturing attention at local, meso, and global scales, our approach offers a versatile tool for comparing relative bias across languages, user subgroups, platforms, or even historical print corpora. Ultimately, our cartographic approach reveals a troubling asymmetry: while publicly accessible data allows for an approximation of global attention patterns, the complete and unfiltered view remains largely hidden behind the closed, proprietary algorithms of major social media platforms, which grant their owners a far more comprehensive view of global information flows.

## The Russia-Ukraine war

In February 2014, following months of Maidan protests in Kyiv and their violent suppression, Ukraine's pro-Russian president Viktor Yanukovych was ousted. Shortly thereafter, on 27 February, Russian forces seized control of Crimea, formally annexing the peninsula in March 2014. This marked the beginning of a prolonged conflict in eastern Ukraine, which has persisted despite the Minsk I (September 2014) and Minsk II (February 2015) ceasefire agreements. Notably, on 24 February 2022, Russia further launched a full-scale invasion of Ukraine with close to 2 million casualties to date by estimations, and numerous human rights violations [1–3]. The war has produced far-reaching global consequences, including disruptions to Ukrainian grain exports to Africa, fluctuations in global oil prices, and broader geopolitical shifts [4]. These global effects underscore the need for large-scale investigations into global attention patterns toward Ukraine.

The Russia–Ukraine war has been characterized as the first fully "social-media war" [5, 6] resulting in an unprecedented volume of content across platforms; although social media has also played a notable role in conflicts before [7]. Governments, militaries, journalists, civilian eyewitnesses, and professional war correspondents have used these platforms for near real-time reporting, while state and nonstate actors deploy strategic narratives, ranging from factual updates to misleading conspiracy theories, to influence public opinion [8–10]. Automated bots further amplify or distort these messages, extending their reach [9, 11, 12].

A rich body of computational social-science research has examined different facets of this conflict across different platforms where Twitter was one of the most dominant sources due to its open API data availability until summer 2023. A major topic of research has been the effects of misinformation and propaganda analyzing the spread of fake news and conspiracy narratives [13–20]. Bot-detection efforts have revealed how automated accounts have skewed attention and engagement in relation to the conflict [9, 11, 12, 21–23]. Other work, for example, focuses on migration flows [24–26], hate speech dynamics [27, 28], polarization and identity formation [21, 29–31], economic and environmental discourse [32, 33], and event-centered topics such as the downing of Malaysia Airlines Flight MH17 [14, 20, 34, 35]. Computational methods involved a range from sentiment or stance detection [26, 32, 36–39] and emotion classification [40] to topic modeling [16, 35], network analysis [30, 34, 41], and geolocation inference [24, 25].

However, most existing investigations remain constrained by narrow temporal scopes, typically focusing on the weeks or months surrounding the 2014 or 2022 invasions, and a limited linguistic range, primarily covering English, Russian, or Ukrainian. This pattern reflects the scope of publicly available datasets, which often span only brief periods around the 2022 invasion, ranging from several weeks to one or two years [42–47]. The studies that do cover a decade mostly limit the scope in some other way, e.g. by only looking at Russian and Ukrainian-language posts, geolocated within Ukraine [29], or restricted to content from specific sources like Bellingcat news [36]. Some studies examine the effects of the war at larger scale by other means, e.g. economic data [48], not to mention a cornucopia of non-computational approaches, not covered in this study. In sum, while there is a large number and broad variety of studies, the broader cross-linguistic and longitudinal patterns of attention towards Ukraine remain largely unexplored. In particular, systematic, data-driven measurements of "attention" or "interest" toward Ukraine spanning multiple languages and a time period of over a decade are notably absent.

To address this gap, we introduce a DNA-microarray–inspired log over-expression cartography that traces added attention to the keyword "Ukraine" on Twitter in 28 languages from 2008 through 2023. By comparing each language's observed mentions against its long-term baseline frequency, our method highlights both prominent spikes corresponding to the 2014 and 2022 invasions, and subtler, language-specific patterns. The resulting macro-scale visualization captures attention dynamics at multiple levels. Local events in individual languages, meso events across languages, and global patterns that allow comparison of different events across time. Notably, our visualization reveals that the two Russian invasions spark attention in two nearly disjoint clusters of languages.

## N-grams

Tracking the frequency of individual words across large corpora can reveal macroscopic cultural dynamics. The concept of n-grams refers to sequences of consecutive words in a text, where n denotes the number of words in the phrase, segmented by spaces. For example, "Ukraine" is a 1-gram, "Volodymyr Zelenskyy" a 2-gram, while "Nothing about Ukraine without Ukraine" is a 5-gram. While the concept has precedent in biology (e.g. k-mers [49]) and linguistics [50], n-gram analysis gained global prominence with Michel et al.'s study [51], which examined the frequency of n-grams in about 4% of all books ever published using the Google Books corpus and assumed that n-grams can be used to detect macroscopic cultural trends, constituting a kind of "culturomics" inspired by the -omics of biology [52].

Our approach, which uses n-grams to detect macroscopic cultural trends, builds on prior research harnessing Google n-grams [51], extended from books to news by the GDELT project [53], and further developed in more recent and robust approaches, subsequently focusing on log changes in Twitter n-gram frequencies [54]. We use the Storywrangler corpus [55] while remaining critical of Twitter's representativeness of culture as a whole, including the sampling and data-balance issues identified for the Google n-grams corpus [56]. As with research into Google n-grams, we confine our analysis to a single, but nonetheless important, cultural "organism" which in our case is Twitter.

N-gram analysis typically involves counting absolute, relative, or normalized frequencies before any semantic interpretation. This simplicity makes it computationally lightweight and well suited for large-scale or real-time monitoring. Indeed, n-grams have been used to track real-time shifts in tweets, including the change of public emotion [57], or to gauge economic uncertainty [58].

However, since Twitter's API access was strategically restricted with Elon Musk's takeover, live trend analyses of public speech on Twitter have been substantially constrained.

Despite these restrictions, extensive historical data remains available. For example, the Storywrangler corpus [55] provides a high-throughput archive of preprocessed Twitter n-grams sampled via the Decahose API (roughly 10 % of all tweets). In this study, we leverage Storywrangler to demonstrate that even a selected subset of n-grams can illuminate global attention dynamics. Having access to this historic archive of public interest, we are afforded a macroscopic view of change in attention towards Ukraine in 28 different languages, covering the full history of the Twitter platform from 2008 to 2023.

## Attention and turbulence

We measure the over-expression of "Ukraine", relative to all of the 1-grams in the corpus, as a proxy for attention. This parallels the "fame" metric [54], where elevated interest is defined as positive frequency in relation to mean expectation. Beyond visualizing rises and falls, our framework enables the quantitative detection of robust patterns: systematic shifts between baseline levels, onset and decay envelopes, exogenous versus endogenous shocks, rank stability, and turbulence [54, 59–61].

Here, we interpret turbulence not as general political instability [62] but as volatility in the frequency of a political term. We focus specifically on turbulence driven by exogenous shocks, events mainly related to war, rather than self-organizing dynamics of a system itself [60]. By contrasting each language's attention level against the expected mean levels of attention, we reveal a spectrum of residual patterns with varying onset sharpness and decay velocities, moving beyond binary classification of shocks [60].

Our approach extends prior work on Twitter n-gram shocks and turbulence [54, 61, 63] and aligns with computational-history frameworks such as cliodynamics [64]. Unlike models that assume fixed cycles or periodicities, we remain agnostic to general rhythms and instead emphasize the emergence of complex, durable patterns from local activity, allowing this method to capture unique events with a characteristic "moyenne durée" shape of mid-term duration [65, 66], longer than a trigger-like event yet shorter than a systematic trend.

## Matrix visualization

Our method transforms plain n-gram frequencies using logarithmic scaling and normalization, displaying them in a matrix format, inspired by gene-expression heatmaps in systems biology [67] and by trajectories depicting deviations from exponential growth in cultural history [66]. Drawing on DNA microarray analysis [68], each cell in our microarray-like matrix (see Figs. 1 and 2) encodes the activation of attention to "Ukraine" in a specific language and time. This matrix representation supports both visual inspection and analytic reordering. By applying semi-algorithmic matrix reordering [69], we reveal coherent clusters and attention gradients that would remain hidden to even educated audiences who are limited to any single language. In this way, our visualization uncovers novel, large-scale patterns of global attention.

## Data

We chose a dataset allowing comparability of attention across time and languages. Our analysis relies on Twitter (now X.com), which offered free or low-cost API access until mid-2023, when costly paywalls dramatically limited access to high-volume data streams for academic purposes. Our data is based on archived materials from the original data-streams, specifically Decahose data, which covers around 10% of all tweets. N-gram frequencies were acquired via the publicly available API at Storywrangling.org [55], making use of the available time frame from 2008 until the cutoff date in Summer 2023, representing uninterrupted coverage of Twitter from 2008 through the mid-2023 transformation of its data-access regime.

We focus on a single, high-signal term — "Ukraine" — in nominative singular to trace shifts in global attention across languages maintaining analytical simplicity (see Supplementary Table S1 for each query). This approach parallels gene-expression studies, where tracking a single well-chosen marker can reveal system-level dynamics; it may hold different functions in different cells, but remains meaningful within each. The choice of a single, prominent term is especially meaningful given the heavy tailed distribution of word frequencies; in the context of Ukraine-related discourse, "Ukraine" is expected to appear with high probability.

We validated that the nominative form of "Ukraine", primarily referring to the country, outperforms related nouns and adjectives in summed relative frequency, with the exception of Hungarian and Finnish (see

Supplementary Fig. S7). In languages where it applies, we chose the capitalized form, but included non-capitalized forms for comparison (Storywrangler treats n-grams as case-sensitive). From the top 44 languages by tweet volume [70, Fig. 2, col. 1], we selected 28 languages supported by Storywrangler, excluding Thai, Chinese and other frequent languages not fully supported by the platform.

## Method

Our analytical approach centers on a simple yet powerful measure of attention across languages and time, to track over- and under-expression relative to given forms of expectation. Our method allows for intuitive comparative analysis via the projection of individual language measurements into a reorderable matrix visualization. This functionally resembles and combines established methods from the areas of DNA microarray analysis [68, 71], the semiology of graphics [67, 69], and computational social science focusing on fundamental patterns [60, 72] within the so-called economy of attention [73]. Similar established plot constructions routinely indicate gene expression in systems biology [74], and have been previously adapted to reveal and track emergent and meaningful fluctuations around the exponential growth of sustained cultural trends on the order of centuries in cultural history [66]. Our perhaps deceptively simple instrumental set-up functions to serve broad audiences, helping to intuitively discern relative attention differences within and across social media platforms, with the power to indicate platform, audience, and language biases, and, following careful consideration, a working proxy for general added excess attention.

### Log-deviation from expected attention

Put in a nutshell, in Figures 1 & 2, we look at the log-deviation from expected weekly log-frequency of the term “Ukraine”, and its equivalents across 28 languages, above and below two baselines of mean frequency expectation: within a given language and across languages. Log-frequency of the term “Ukraine”, to be precise, refers to the fraction of mentions of “Ukraine” in relation to all other 1-grams on a logarithmic scale. We use this relative measure because raw counts in the Storywrangler corpus, based on the roughly 10% Twitter Decahose sample, depend on overall Twitter activity, which varies strongly across time and languages. The measure therefore captures the share of discourse devoted to "Ukraine" rather than the absolute number of mentions.

The matrix visualizations of the resulting temporal trajectories of attention dynamics, as exemplified in Figures 1 & 2, strikingly reveal emergent resonance and differences in relative impact across languages.

In Figure 1 individual language trajectories are independent, simply indicating the deviation of attention from the mean expectation within a given language over time. Figure 2 is renormalized to a different baseline, indicating the relative deviation from the expected attention “market share” across all languages.

For this purpose, the mean frequency of “Ukraine” across time within a given language is considered to be the expected frequency of the language. The expected “market share” of the language is the relation of this expected frequency to other languages. In both cases, Figures 1 & 2, diverging from the respective attention baseline is considered over/under-expression, reflected by the colors red/blue.

For clarity, in both Figure 1 & 2, each matrix cell element indicates two values, the total attention in a given moment, indicated by element bar size, and the deviation from the baseline, indicated by color. The element bar size in each matrix cell indicates the relative share of attention in a given week and language, i.e., the fraction of mentions of “Ukraine” in relation to all other 1-grams, on a logarithmic scale. Meanwhile the color indicates the log deviation from the expected attention.

We provide the matrices of expected attention within each language and between all of the languages. These two types of attention divergence mean that Figure 1 shows the surprise-like effect in each language and Figure 2 shows which of the languages have more (over-expression) or less attention in a given week in relation to other languages and in average across time. Harnessing the fact that reordering the heatmap matrix does not change the inherent cell values, we reorder the matrix rows semi-algorithmically, following Bertin [69], to reveal meaningful clusters and gradients. To approximate the news-cycle decay [75], we smooth the within-language series over three weeks and the market-share series over six weeks and provide non-smoothed versions as supplementary material (Fig. S2 and S3).

The results of this matrix become especially salient through semi-algorithmic matrix reordering [69], clustering languages with similar attention trajectories into contiguous blocks. This uncovers gradients and patterns beyond the cognitive reach of any single-language viewer. A comparison with a raw log-frequency heatmap (Supplementary Fig. S1) clarifies the advantage of our approach: while raw log frequency indicates global patterns and resonance, the additional lens of log under- versus over expression reveals salient events with greater clarity. Our case study, utilizing such an interdisciplinarily inspired method of graphical construction [66, 67, 69], brings salient wartime events and language groupings into sharper focus.

### Clustering resonant languages and periods of time

To uncover historic resonances in attention dynamics, we order and cluster both languages and periods of time. In Figures 1 and 2, languages are initially arranged manually to reflect coherent groupings, contrasted with hierarchical clustering using Ward's linkage method (see dendrogram in Figure S5).

For a systematic, period-by-period comparison, we construct a 28-dimensional attention vector for each six-week interval, where each element represents the summed relative frequency of 'Ukraine' in one language over these six weeks, mirroring the moving windows used in the heatmaps. We project these vectors into two dimensions with t-SNE (perplexity = 15, Euclidean metric), revealing clusters of similar global attention profiles across the entire 2008–2023 span (Figure 5a). This projection makes it possible to identify periods that share similar cross-linguistic attention patterns even when separated by years and interspersed with unrelated events. Complementarily, Figure 5b presents a Jensen–Shannon distance matrix of the same vectors (the square root of the Jensen–Shannon divergence), akin to gene-expression distance arrays, mapping shifts and resonances throughout the fifteen-year period. Distances reflect the relative distribution of attention across languages rather than the overall volume of attention, and are shown on a logarithmic scale.

Finally, to dissect the fine-grained dynamics around the two major escalations, the Russian invasions in 2014 and 2022, we clustered the onset-and-decay patterns of attention in different languages around these periods. We normalize each language's daily series by its mean within the clustering window and apply K-means clustering to the matrix of pairwise Dynamic Time Warping distances between languages [76]. For 2022, both clustering and plotting use the full eight weeks (−4 to +4w) centred on 24 February. For 2014, we cluster on a narrower four-week window (−1 to +3w) centred on 18 February, the first and deadliest shock, but plot it within the full −4/+4w context in Fig. 3; Fig. 4 instead aligns on 27 February to match the Crimea seizure as the invasion date proper. Window lengths, cluster counts, and the clustering algorithm were chosen qualitatively to best capture the shocks, onset, and offset of these events; the cluster metrics in Supplementary Fig. S4 show no clear optimum and served only as a secondary check. Figures 3 and 4 present these clustered time series, illustrating distinct onset and decay profiles that exemplify meso-level patterns of attention.

## Results

We report findings across three analytical scales for the sake of clarity: micro scale, capturing specific events unique for some languages; meso scale, examining particular event related time periods across languages; and macro scale, revealing the clusters of languages and different events across time. Figures 1 and 2 present a cartography of attention dynamics toward "Ukraine" on Twitter over a 15-year period, across 28 languages, offering visual input for comparisons at all scales. Figure 3 reports meso-scale findings by clustering the languages' response profiles within each invasion period separately, while Figures 4 and 5 display macro-level comparisons across events, first between the 2014 and 2022 invasions and then across the full 2008–2023 period. We report results for dominant events exhibiting marked overexpression.

The visual language we employ constitutes a methodological contribution to the interpretation of large-scale historical data. Figures 1 and 2 demonstrate the utility of a DNA microarray-inspired matrix approach for capturing historical changes comparatively and at scale. In addition, we incorporate an economics-inspired "market share" visualization to accentuate specific shifts in attention. Figures 3 and 4 relate to the framework of endogenous and exogenous shocks introduced by Crane and Sornette [60], illustrating the clustering of varied responses to the major external shocks of 2014 and 2022. Unlike Crane and Sornette's discrete classification model, our approach adopts a data-driven classification describing a spectrum of patterns with varying onset sharpness and decay velocities, better accommodating the complexity and heterogeneity of the real-world events under investigation here. Finally, Figure 5 facilitates a global-scale vector-driven comparison of events across time again in the style of DNA microarray analysis.

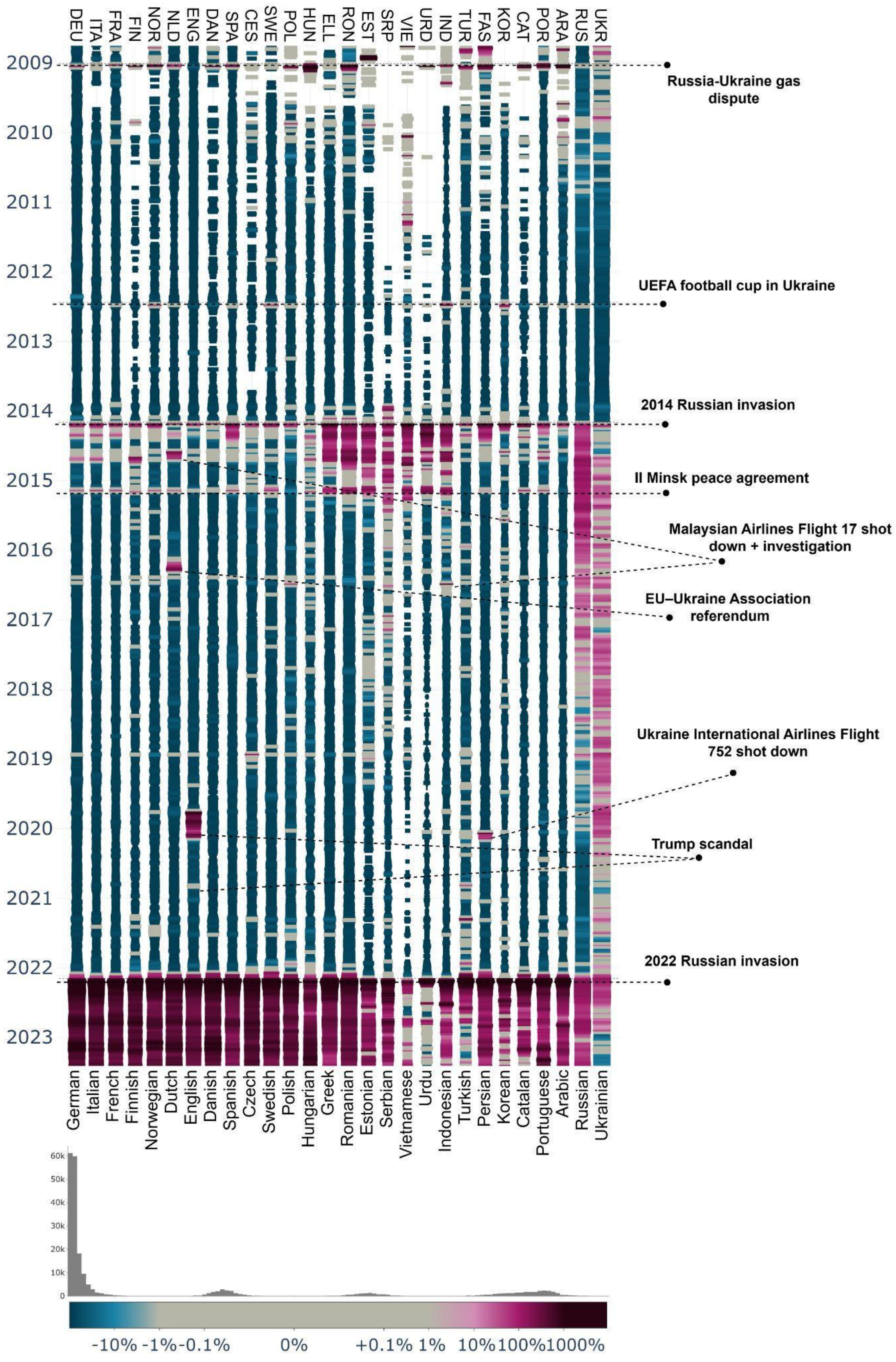

**Fig. 1. Log over/under-expression from expected log frequency within languages exhibits dramatic changes in 2014 and 2022 in all languages.** Each trajectory corresponds to mentions of "Ukraine" in one of 28 languages across time from the end of 2008 until summer 2023. The bar size of each data point shows the log-scaled sum of mentions in relation to all other 1-grams that week. Color indicates the deviation from the expected frequency, which in this case is the average mentions in the same language across time. Red hues mark weeks of high attention, blue the weeks of low attention and gray tones close to average attention. A three-week backward-looking moving average is applied to attention deviation to reduce noise and account for assumed audience memory effects.

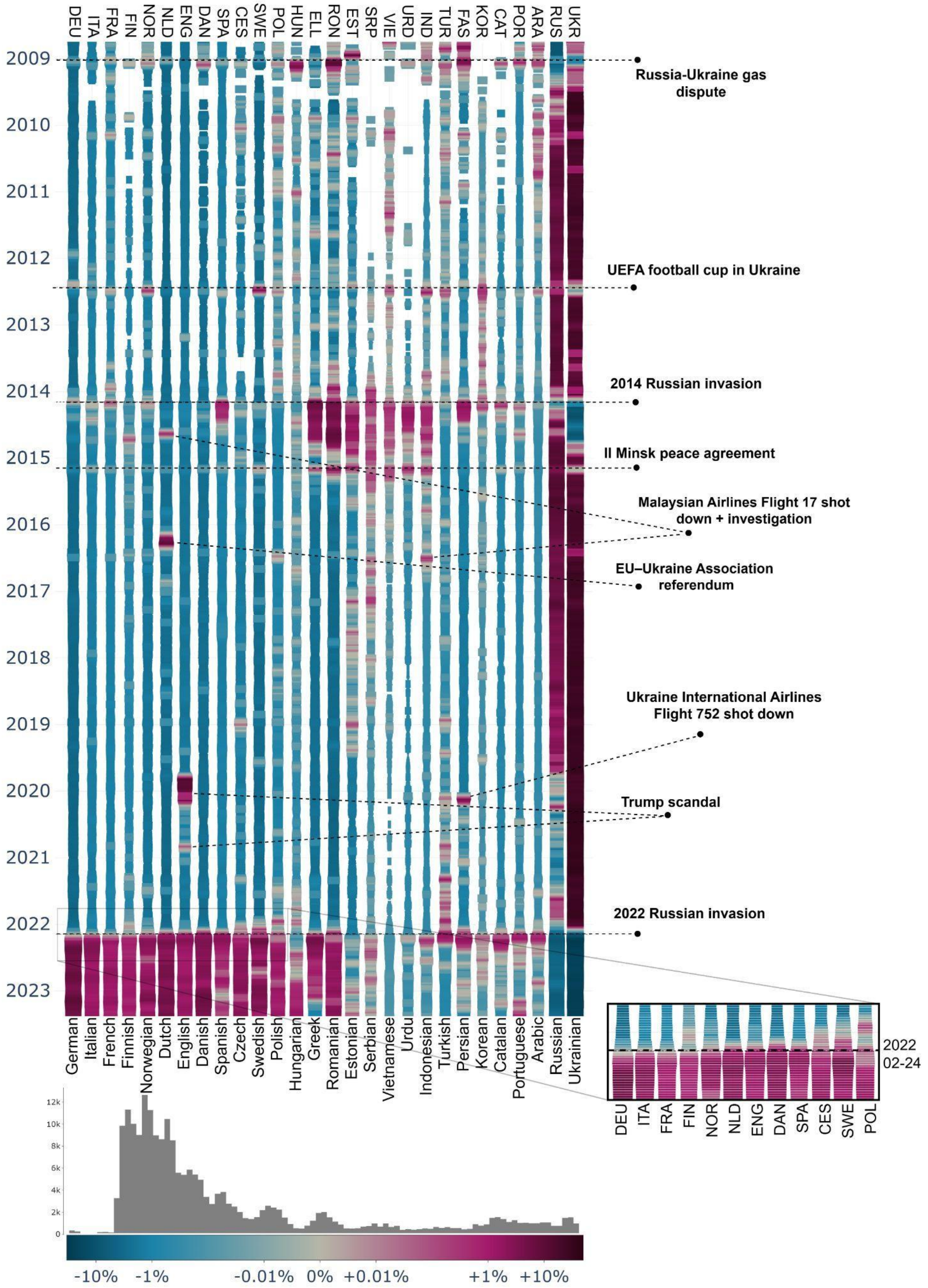

**Fig. 2. Log over/under-expression from expected log frequency shows language groupings and rankings of overexpression.** Unlike the within-language deviations in Figure 1, the colors in the matrix here show each language's weekly attention deviation from the global expected "market share" of attention. The smoothing is over six weeks. Language trajectories are arranged manually, informed by hierarchical clustering (Fig. S5), so that similar languages sit adjacent, revealing coherent groupings and their collective attention patterns. Note that relative market share of language is interdependent: Ukrainian and Russian attention, for example, appear very low during invasion times, due to the relative increase in attention market share over the expectation in other languages (compare with Figure 1).

## Micro scale events are contextualized across languages

First, we highlight individual events in specific languages that distinctively stand out over time in Figures 1 and 2. Although these events are well known, they have not previously been mapped across languages and time or compared in terms of relative attention in a data-driven framework.

- On 17 July 2014, the downing of Malaysia Airlines Flight MH17 triggered a pronounced over-expression in Dutch, reflecting the large number of Dutch victims, though several other languages also registered noticeable reactions.
- A second Dutch-language surge occurred in January–April 2016, as public debate intensified ahead of the 6 April referendum on the EU–Ukraine Association Agreement.
- In September 2016, the Joint Investigation Team's presentation of its criminal investigation findings coincided with an over-expression in Indonesian, likely driven by the Malaysia airline, passengers, and crew members involved in the crash.
- The 2019 Trump–Ukraine scandal, when President Trump withheld U.S. aid and pressed President Zelensky to investigate Joe Biden, elicited a prominent spike in English, which then waned as attention shifted toward COVID-19.
- the January 2020 shoot-down of Ukraine International Airlines Flight PS752 over Iran, with a Ukrainian crew and mostly Iranian and Canadian passengers, produced a marked increase of mentions in Persian, followed by a rise in Turkish mentions in February 2020, perhaps reflecting President Zelensky's subsequent visit to Turkey.

These language-specific patterns underscore how singular events can generate distinct attention dynamics that are best understood when compared across linguistic communities and time.

## Meso scale patterns highlight language differences across invasions

### *Global shocks*

Beyond events in single languages, we identified the five most notable events with strong added attention spanning across many or all languages (see Figure 1). The earliest, trigger-like event, was the Ukraine-Russia gas dispute at the end of 2008 and the beginning of 2009, affecting gas availability across Europe. This event remains visible despite the relative scarcity of Twitter content at the time and is accentuated by smoothing (compare to Fig. 2 and Supplementary Fig. S3). In 2012, non-war-related event of the UEFA European Football Championship, partly hosted in Ukraine, elicited increased attention across many of the languages under study, although to a lesser extent than the later conflict-related events. In 2014 global attention peaked around the Maidan protests, ousting of pro-Russian Ukrainian president and Russian annexation of Crimea. The February 2015 Second Minsk peace agreement also generated 10%-100% overexpression in many of the languages. By far the strongest peak is the 2022 Russian invasion, which dominates the map (Figures 1 and 2) with overexpression in all 28 languages (Fig. 1), persisting well into 2023 in all but Ukrainian, Turkish and Urdu.

### *2022 and 2014 invasion clusters*

The relative magnitude and timing of attention spikes during the 2014 and 2022 invasions reveal both commonalities and divergences across languages. As shown in Figure 1 and accentuated in Figure 2, every language exhibits some reaction, but several, Greek, Romanian, Estonian, Serbian, Vietnamese, Urdu and Indonesian, sustain elevated attention well beyond the initial shock in 2014. These same languages also respond notably to the Second Minsk Agreement, implying prolonged engagement from the Maidan events onward. Persian and Spanish display a moderate, extended response with a smaller spike at the Second Minsk Agreement.

The 2022 dynamics reveal Vietnamese and Urdu showing the smallest and shortest-lived shocks (Fig. 1), whereas the "market-share" based over-expression in Figure 2 highlights a pronounced, sustained shift in a group of European languages — German, Italian, French, Finnish, Norwegian, Dutch, English, Danish, Spanish, Czech, Swedish, Polish, Greek, and Romanian. Hungarian is an exception with a smaller shock than other European languages, but with higher than expected sustained attention. By contrast, Russian and Ukrainian remain high-attention outliers, alongside a mostly non-European group (Turkish, Persian, Korean, Portuguese, and Arabic, joined by Catalan), whose initial shock subsided into a moderate but still persistent interest. This pattern likely reflects Europe's enduring focus on the conflict versus a more transient attention elsewhere.

A somewhat striking finding is revealed in the overexpression dynamics of the same market-share trajectories, if we zoom into detail: During the 2022 invasion, several European languages crossed the market share threshold to high-signal attention at different times: Swedish and Polish react the earliest, followed by Czech, Spanish, Danish, English, Dutch, Norwegian, and Finnish, with the French, Italian, and German reaction essentially looking like a sudden first-order transition from non-attention to shocking alertness. This ordering intuitively reflects known shifts in the dominant narratives of some languages — for example, the disillusionment and alteration of formerly pro-Russian sentiments in Germany, Finland, and Italy after the invasion — calling for further comparative studies of specific language and national dynamics governing the global complexities of sentiment at that time.

To probe these dynamics further, we applied time-series clustering around each invasion, identifying five clusters in 2014 and six in 2022 (Figure 3). In 2014, two principal shocks emerge: the first around 18–20 February, coinciding with the deadliest Maidan protests, and the second on 1–3 March, following the Russian parliament's authorization of use of force in Ukraine. Notably, the 27 February occupation of Crimea produced only a modest immediate spike but fueled sustained attention in specific languages. By contrast, the 2022 invasion generated a single, dominant shock on 24 February, which yields clearer distinctions among clusters based on shock intensity, duration, and decay.

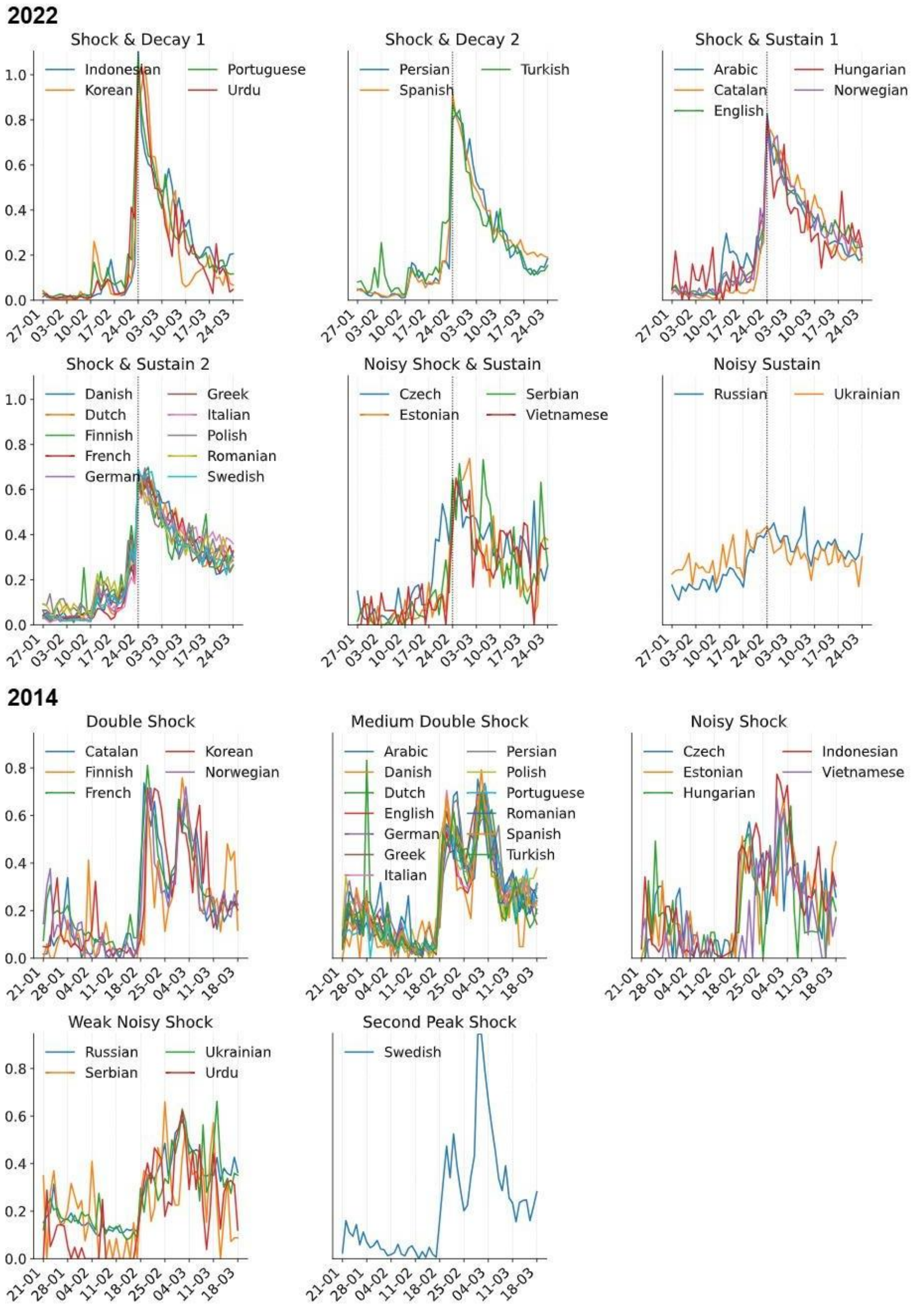

**Fig. 3. Time-series clusters for 2014 and 2022 invasions show a complex typology of attention dynamics across languages.** Each subplot is one cluster, each line one language, normalized by its mean over the visible 8-week window (log-transformed). Both 2014 and 2022 are plotted 4 weeks before and after 18 February 2014 and 24 February 2022 (dotted line), but clustered on −1 to +3 weeks for 2014 and full −4 to +4 weeks for 2022. No anchor line is drawn for 2014, where attention followed several successive shocks rather than one event. See Supplementary Fig. S4 for cluster evaluations.

2022 invasion clusters (ordered from sharpest shocks to most sustained response):

- *Shock & decay 1* – sharpest shocks followed by rapid decay in Indonesian, Korean, Portuguese and Urdu.
- *Shock & decay 2* – sharp shock but with a transition to a higher steady attention level than in the previous cluster. Includes Persian, Turkish and Spanish.
- *Shock & sustain 1* – more moderate shock with more prolonged decay and an elevated new baseline in this context window for Arabic, Catalan, English, Hungarian and Norwegian. Arabic patterning with these otherwise European languages may hold several reasons: the Arab diaspora in Europe, actual similar attention in Arab-speaking countries, or other reasons not identified here, including bot activity.
- *Shock & sustain 2* – smaller shock with some sustain on the top of the shock indicating less surprise but a more sustained attention. Decays to a higher baseline than previous clusters. Notably all of the 10 languages, Danish, Dutch, Finnish, French, German, Greek, Italian, Polish, Romanian, Swedish, are European and widely spoken in countries that continued the support and attention towards Ukraine; with notably different attention dynamics than e.g. Indonesian, Portuguese, Korean or Urdu.
- *Noisy shock & sustain* – weaker over-expression and noisier signals characterizing Czech, Serbian, Estonian and Vietnamese. Czech shows similarity to other European languages (*Shock & Sustain 2*) and has pronounced earlier attention.
- *Noisy sustain* – Russian and Ukrainian exhibit only slight increases above their already high baselines, reflecting the centrality of "Ukraine" to its own identity, and also its relevance within Russian. Notably, we see that Ukrainian attention rises beginning 17 February, a pattern absent elsewhere except notably in Czech (cf. *Noisy Shock & Sustain*). The method does not distinguish Russian speakers in Ukraine, bot activity, and similar potential differences in attention.

2014 annexation clusters:

- *Double shock* – strong shocks around both the February 18 Maidan killings and the March 1 Russian parliamentary vote, ending at a slightly higher attention level. Includes largely European languages, like Norwegian, Finnish, French (cf. Limitations), Catalan, but also Korean. Together with the next clusters, they reflect the surprise of the EU in reaction to the violence.
- *Medium double shock* – double-shock pattern with less intensity and smaller drops between the two shocks. Somewhat higher attention around the 1 March. This cluster describes many European languages, but also includes Arabic (similarly to the 2022 invasion), Persian and Turkish.
- *Noisy shock* – weaker initial shock related to Maidan and generally more noise but with a pronounced second peak at the Russian declaration. Similar to the 2022 cluster for languages with less Twitter usage. Includes Czech, Estonian, Hungarian, Vietnamese and Indonesian.
- *Weak noisy shock* – characterized by minimal distinction between the two peaks (no intervening drop) and relatively greater focus on 1 March. This cluster includes Urdu and Serbian, languages that otherwise showed low overall attention to Ukraine, as well as Ukrainian and Russian.
- *Second peak shock* – Swedish shows a pronounced second peak that exceeds all other clusters within the normalized window, while otherwise resembling the *Noisy Shock* cluster.

## Macro scale observation summarizes system-scale historical resonance

The macro scale illustrates how cultural-data analysis can surface patterns that remain hidden when examining individual events or languages in isolation. By comparing distinct moments in time, we can observe how languages cluster across Twitter's history in their attention to "Ukraine" and trace the similarities between periods with different levels of global attention.

*2014 and 2022 comparison*

Our quantitative view shows the highest peaks of Twitter attention to "Ukraine" in February 2014 and February 2022 (Fig. 1), yet the dynamics of these moments diverge markedly (Fig. 4). Across languages, the 2022 peak exceeded the 2014 peak by a median factor of 8.5, and by at least an order of magnitude in 13 of the 28 languages, reaching 44× in Portuguese and 20× in Arabic. Russian and Ukrainian exhibit the smallest relative increase, maintaining a consistently high baseline, while Vietnamese stands out with a lower peak in 2022 than in 2014, likely reflecting a sizable Vietnamese community within Ukraine. Romanian and Greek also demonstrate sustained high attention in both periods, with 2022 peaks only three to four times higher than in 2014. Moreover, attention in 2022 culminates in a single, dominant surge on 24 February, preceded by a gradual rise during the pre-invasion troop build-up, whereas the 2014 surge unfolds in two to three distinct subpeaks (except in Czech and Vietnamese) related to the protests and the invasion.

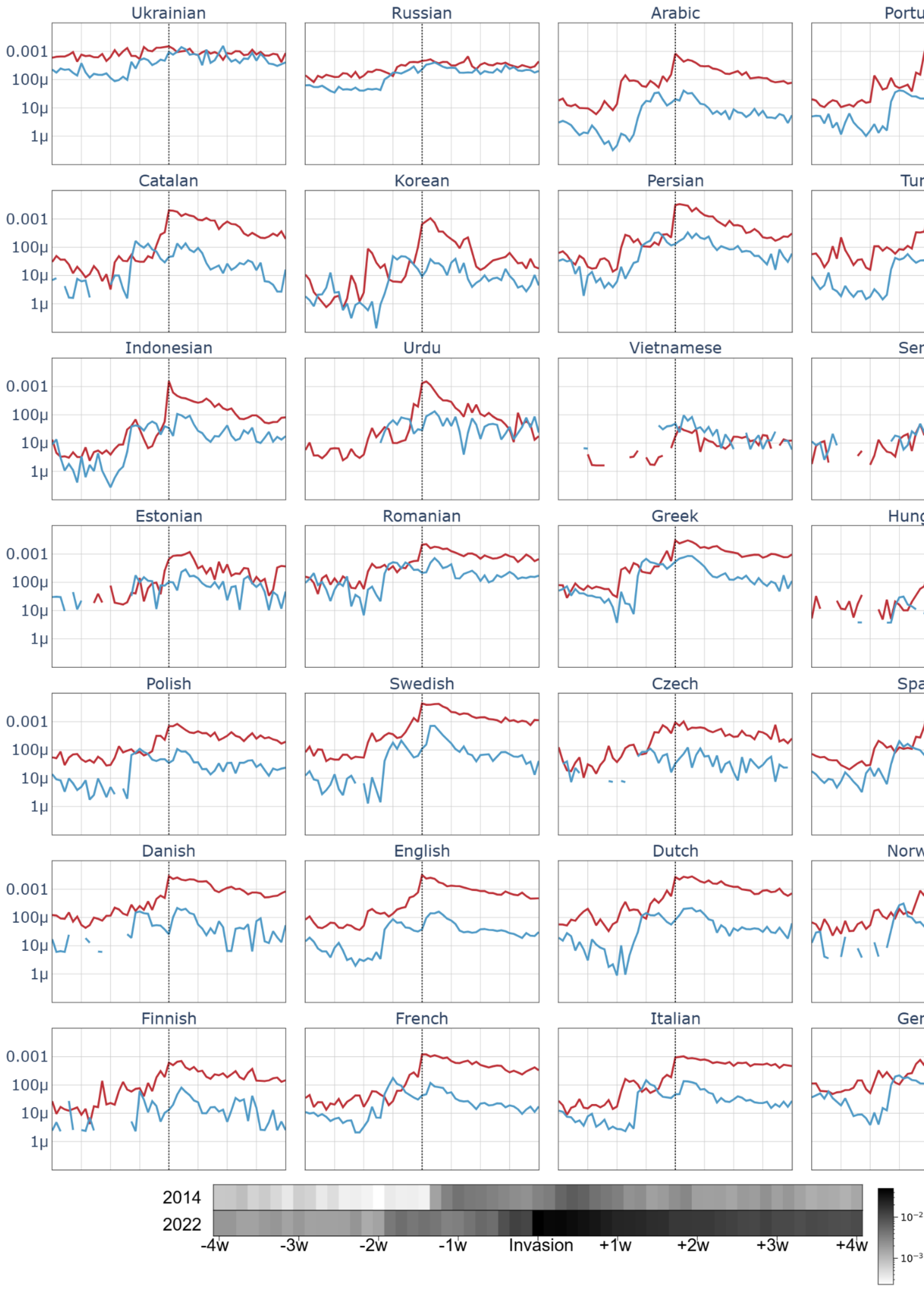

**Fig. 4. Pairwise comparisons of 2014 vs. 2022 invasion periods indicate markedly different dynamics.** Each subplot shows attention trends for 2014 (blue) and 2022 (red) from four weeks before to four weeks after the invasion date per language. The vertical line marks 27 February 2014 and 24 February 2022. The logarithmic y-axis underscores deviations spanning orders of magnitude. The heatmap strips below display the summed frequencies for each year. Distinct onset and decay envelopes emerge, with most languages sustaining elevated attention after the 2022 invasion, far more than in 2014. Notably, in several languages the 2014 and 2022 envelopes share a similar shape despite the different triggering events, as if each linguistic community acted as a resonant system whose own properties shape the onset, sustain, and decay of attention, largely independent of the specific input.

### *General language clusters*

Strikingly, the 2014 and 2022 attention dynamics *separate* our 28 languages into five core *groups* plus outliers, based on the temporal distribution of expressed *attention.* We select*ed* these groups manually, informed by Figs. 1, 2 and S5, and show them in Fig. S6. These groupings emerge most clearly in the "market share" perspective and are also evident when comparing under- and over-expression within each language (see Figs. 1 and 2). The groupings reflect different geopolitical positions.

*Ukrainian and Russian*. These two languages maintain persistently high attention to "Ukraine", reflecting everyday usage, either self-referential or narratives constructing the "other". Their relative market shares only fall when other languages surge, producing an apparent under-expression artifact around 2022 that in fact signals crisis-driven spikes elsewhere and making these languages a relative point of stability with which to compare the larger shocks in other languages.

*Turkish, Portuguese, Arabic, Catalan, Korean*. Together with Persian and Indonesian as outliers, this transcontinental group shows the most short-lived reactions to both invasions. The Portuguese dynamics may suggest divergent political stances in EU-Portugal versus Brazil, as their separate attention could differ (studying Tweets including their geolocation is out of scope of our dataset; yet cf. [77]). Turkish attention was sustained post-2022, but it exhibits multiple over-expressions between 2014 and 2022, hinting at a more general engagement before the conflict.

*Estonian, Serbian, Hungarian, Vietnamese, Urdu*. These languages, with Persian and Indonesian as partial outliers, exhibit sustained added attention in 2014 but not in 2022. Their lower overall Twitter volumes may amplify relative changes. Estonia and Serbia, in particular, spike even when most other languages do not, mirroring Turkey's pattern but earlier in the conflict. Notably, Serbia and Hungary have historically leaned more pro-Russian than Estonia, while Vietnamese and Urdu are geographically distant from the conflict. Hungarian also shows a late uptick toward the end of the observation period, perhaps reflecting shifting domestic discourse.

*German, Italian, French, Finnish, Norwegian, Dutch, English, Danish, Czech, Swedish and Polish*. With Spanish as an outlier with more attention in 2014 and a faster decline after the 2022 shock, these predominantly EU-spoken languages show limited reaction in 2014 but strong, prolonged attention in 2022 (in comparison, Hungary characteristically gains that attention later in time). This enduring over-expression reflects both the war's heightened relevance for European audiences and a narrative shift from the muted response in 2014 toward sustained engagement in 2022.

*Romanian and Greek*. Both stand apart from other European languages with strong, sustained over-expressions in 2014 and 2022. Additionally, Romanian exhibits an early spike during the 2009 gas-price crisis, alongside Hungarian. It is noteworthy that Greece [78], Romania and largely Romanian speaking Moldova have all been dependent on Russian gas, underscoring their geographic and economic ties to Ukraine and Russia. The increased attention in Romanian is also expected considering the political effect of the conflict in the long term [79].

### *Event similarities and complexities*

To obtain a clearer macro-scale view of temporal affinities beyond the two invasion peaks, we represented every six-week window from 2008 to 2023 as a 28-dimensional attention vector and mapped these into both a 2D t-SNE projection (Fig. 5a) and a Jensen–Shannon distance matrix (Fig. 5b). Attention over time is far from uniform: instead of evenly spaced intervals, we observe tight clusters, e.g. post-2022 invasion (Fig. 5a). Notably, the 2022 invasion profile is most similar to other globally high-attention periods, such as the 2009 gas crisis, UEFA 2012 football championship, and the 2014 invasion (Fig. 5b), suggesting that structurally similar cross-linguistic attention profiles can recur across events that differ vastly in scale and context. By contrast, the 2014 invasion appears as a sudden departure from preceding activation patterns but does not coalesce into a large cluster, reflecting its shorter sustain envelope and greater variation across languages.

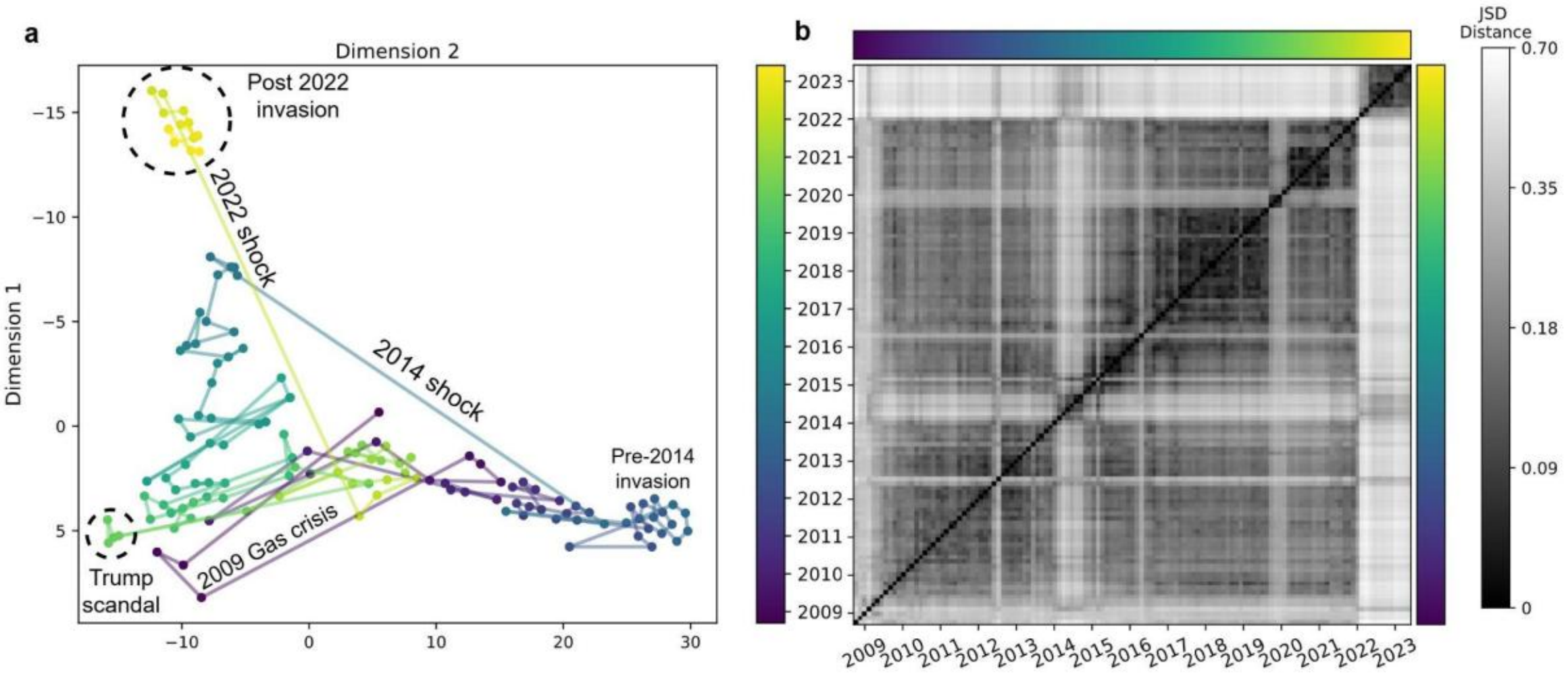

**Fig. 5. Date similarity comparison reveals clusters of similar dates related to different levels of global attention.** t-SNE projection and Jensen–Shannon distance matrix based on vectors of six-week summed relative frequency (share of total frequency per week), where each language is one dimension. a) Color coding in the t-SNE projection ranges from dark blue (2009) to bright yellow (2023), whereas similarly colored links between the points represent linear timeflow. b) Jensen–Shannon distances between the same vectors, shown on a logarithmic color scale; darker tones indicate the most similar (smallest-distance) six-week periods. The color bars along the top and right edges encode time using the same color scale as in a).

In sum, we have revealed cross-linguistic attention patterns that would elude any monolingual perspective, offering historians and media scholars quantitative reference points that are grounded in coherent quantification combined with visual analysis. The invasion events are aggregated, but comparing local, meso, and macro level allows to balance between a general overview and a more in depth complexity of an "event", such as the 2014 invasion unfolding through a series of nested shocks from Maidan protests to Crimea's annexation.

At the micro scale, Malaysian and Iranian airlines disasters and the 2019 Trump scandal register as the most pronounced language-specific spikes. At the meso scale, five global events, the 2009 gas crisis, the UEFA 2012 football event, the Second Minsk Agreement, and the 2014 and 2022 invasions dominate attention clusters, with the 2014 surge displaying richer sub-peak structure than the sudden single-peak 2022 invasion.

Macro scale is where the most consequential and otherwise hidden patterns appear. Here, every language shows some reaction to the identified events, but "market-share" deviations and clustering distinguish nuanced response groups: many European languages mount weaker responses to the 2014 crisis than expected, yet sustain heightened attention for 2022; anticipation and reaction delays vary from German to Swedish; and only Romanian and Greek combine shock with sustained attention in both periods. Non-European languages, such as Korean, Persian or Arabic, follow dynamics related to European languages, but with marked differences, especially for the 2022 crisis, underscoring the regional relevance.

Ultimately, our main contribution lies in providing a quantitative cartography that supports and deepens qualitative expert interpretation. By tracing attention to Ukraine across languages and time on Twitter, we show how events produce distinct reaction profiles with differences and overlaps uncovering new and meaningful patterns in global attention.

# Discussion

Our study fills a gap in the literature by offering a large-scale, data-driven overview of attention to the Ukraine–Russia war, alongside a novel interdisciplinary methodology that integrates time-series sampling dynamics, DNA microarray analysis [68], the semiology of graphics [67, 69] and useful terminology from study of economic complexity. The resulting cartography of historical attention can provide historians, media scholars, public-sector stakeholders, and a broader audience with quantitative reference points grounded in visual analytics, enabling a broad perspective on politically significant events and crises. We also recognize the limitations of this study: reliance on a single keyword, and linguistic and sampling constraints of the dataset, that both warrant future expansion and validation.

### Keyword validation

We selected the nominative form of "Ukraine" in each language to maximize signal strength. Comparison with related nouns and adjectives (see Supplementary Fig. S7) confirms that this form yields the highest usage in all languages except Hungarian and Finnish, where a genitive/adjectival form is more frequent; we retain the nominative for cross-language comparability. In languages with rich morphology, e.g. Estonian, dozens of related inflections exist, but only a few appear frequently enough in the Twitter Decahose (10 %) stream. We therefore argue that our chosen form provides a robust first-order approximation, while acknowledging that shifts to variants (e.g., "Ukrainians") or polysemous uses of the nominative may introduce noise. Just as particle traces in fluid dynamics cannot capture full flow complexity, or a single gene's expression in microarrays cannot reflect all cellular functions, our n-gram proxy simplifies a richer discourse dynamic.

## Limitations

The study's linguistic limitations stem from the wide variation in language rules and meanings across different language families, where a concept in one language may be expressed differently or split across multiple words in another. Using 1-grams means capturing only a single word even in analytical languages such as Vietnamese and Indonesian where concepts like English "Ukrainian" are expressed with multiple words.

Our reliance on Twitter data was driven by the platform's accessibility at the time of collection; although the raw dataset remains available, it now represents a historical snapshot and cannot support real-time or live-stream event mapping. For example, the dataset ends before the onset of the Israel–Palestine war, or the escalation including Iran, and the topics surrounding Trump, Russia, and Epstein, which may have shifted and may keep shifting the relative distribution of attention away from Ukraine. We are careful in generalizing beyond the language to underlying socio-political situations, and we cannot verify the geographic or demographic sources of attention: for French, Portuguese, and Spanish, we assume coverage by both European and Latin American speakers, but cannot disaggregate these contributions. The same goes for diasporas and the shifts in migrant communities speaking different languages that this study does not account for. Finally, language classification itself may in some cases be erroneous [55, 70], introducing noise or bias to our dataset. Furthermore, the mix of Twitter contributors, from corporate and institutional accounts (incl. government members and embassies) and news outlets to everyday users, and malicious artificial agents has possibly shifted over time, resulting in complex changes in hidden narrative baselines and goals of the posts.

Moreover, by focusing on n-gram frequencies to track attention dynamics, we were unable to filter out manipulative content. Twitter has served as a major conduit for Russian disinformation [80, 81] and hosts active pro-Russian bot networks [11] that may have skewed our attention metrics. While our qualitative review of popular tweets across different periods of time did identify tweets echoing pro-Russian strategic narratives, a full systematic analysis of disinformation falls beyond the scope of this study.

## Future Directions

There are several avenues to deepen and broaden the application of our framework, and, by application, our understanding of the Russia–Ukraine war, ranging from comparative n-gram studies to disinformation analysis. A comparative n-gram approach could treat terms like "Ukraine" and "Russia" as paired probes, generating heatmap matrices of attention differences that reveal shifts in the representative discourse prominence. Our pilot study indicated that "Russia" dominated pre-2014 conversations but was then overtaken by "Ukraine" after the invasions, suggesting a shift in these identity markers. Second, our method's analogy to microarrays invites a disinformation-vs-organic comparison, contrasting streams known to contain automated or state-sponsored messaging against user-generated content. Quantifying how bot campaigns inflate or distort attention metrics would strengthen our ability to grasp the genuine public interest from coordinated amplification at scale.

This approach can be further strengthened by focusing on a bundle of n-grams instead or by integrating user metadata, such as account type (embassy, news outlet, NGO, individual), follower counts, verification status, and geolocation to clarify which actors drive spikes in each language and how their influence shifts over time. Moreover, the attention cartography contains far more interpretive potential than we have explored here, particularly when examining what does not attract attention, or considering that some topics are suppressed in one social media platform versus another. Our focus here was on the comparative aspect of events that were highlighted in at least one language, even though some events may have gained more attention through specific keywords rather than through "Ukraine".

Further accommodating demographic and platform-evolution factors (e.g., migrant community growth or generational adoption patterns) would enhance the interpretation of the analysis. Yet even with our current "strongest-signal" focus, investigating smaller fluctuations promises some additional insights. Finally, extending the dataset beyond mid-2023 and adapting the toolkit to other platforms, or multi-platform composites, would allow better capture of emerging crises (e.g., the Israel–Palestine war) and the evolving rhythms of public attention.

# Conclusion

Our multi-scale investigation of Twitter attention to "Ukraine", spanning micro, meso, and macro level cartographies sheds light on how global public interest has shifted from 2008 through mid-2023. At the micro scale, distinct tragedies and political controversies stand out: the downing of Malaysia Airlines Flight MH17 in July 2014, Dutch EU-Ukraine referendum of 2016, the Ukrainian passenger plane shot down in Iran in January 2020, and the 2019 Trump–Ukraine scandal, each generating sharp, language-specific spikes in mentions. At the meso scale, five major episodes dominate collective attention: the 2009 Russia–Ukraine gas dispute, UEFA Euro 2012; the 2014 Maidan protests and annexation of Crimea; the Second Minsk peace agreement; and the invasion of 2022. A macro-scale perspective compares the temporal events, revealing distinct dynamics of 2014 events with multiple shock waves and 2022 invasion resulting in intense and prolonged surge across most languages. We illustrate that global attention is not evenly distributed over time but organized into distinct temporal clusters around major events. Moreover, structurally similar attention profiles recur across events separated by years and differing in scale — the history of global attention, it appears, does not repeat but rhymes.

Strikingly, we find coherent clusters that would remain invisible to any single-language audience, uncovering novel, large-scale patterns of global attention. Across all 28 languages, "market-share" deviations (Fig. 2) uncover: predominantly EU languages that lack the strong shock with sustained attention in 2014 in comparison to 2022. Moreover, there is a clear hierarchy in the onset of the 2022 response: Swedish leads, while Italian, French, and German lag with the slowest onsets. Secondly, lower-volume languages — Estonian, Serbian, Vietnamese, Urdu, and Hungarian — exhibited much more notable sustained attention in 2014 and registered modest over-expressions in cases when attention in other languages waned. Thirdly, several languages spoken mostly outside Europe, including Persian, Portuguese, and Arabic, display brief surges in both 2014 and 2022 but without long-term persistence. Lastly, Romanian and Greek stand out with especially strong initial reactions and prolonged attention across both invasion periods. These macro-level patterns highlight how geopolitical proximity, linguistic networks, and regional media ecosystems shape the diffusion, timing, and endurance of global attention.

Methodologically, we demonstrate that adapting techniques from gene-expression heatmaps [67], exponential-growth deviation plots [66], DNA microarray analysis [68], and semi-algorithmic matrix ordering [69] enables the mapping of large-scale attention dynamics in a manner that is both computationally robust and qualitatively interpretable. Like node–link diagrams in network analysis, our cartographic visualizations expose coherent clusters and attention gradients that no single-language view could reveal, offering historians, media scholars, and policy analysts new vantage points for understanding how geopolitical events propagate across linguistic communities and over time. Unlike node-link diagrams, our matrix visualizations are easy to read even for broader audiences, essentially providing a symphonic listener's partiture of cultural history.

Finally, this study highlights the importance and fragility of access to large-scale social-media data. Just as weather forecasting depends on open and timely information, the transparent study of digital public spheres requires publicly accessible data, tools, and infrastructures. Without such access, we risk entrenching asymmetries in who can observe, analyze, and ultimately shape the global flow of information. Our framework offers one step toward a more systematic and interpretable cartography of online attention, but its future potential depends on maintaining and expanding equitable access (cf. EU Regulation 2022/2065, Art. 40 [82]) to the digital traces through which societies understand themselves.

# Abbreviations

API, application programming interface; DNA, deoxyribonucleic acid; EU, European Union; GDELT, Global Database of Events, Language, and Tone; NGO, non-governmental organization; t-SNE, t-distributed stochastic neighbor embedding; UEFA, Union of European Football Associations.

# Declarations

### Availability of data and materials

The n-gram datasets analysed in this study are publicly available via the Storywrangler project (https://storywrangling.org/) [55]. The Python code used for data collection (including API queries) and analysis is publicly available at https://github.com/markmets/Ukraine-Twitter.

### Competing interests

The authors declare no competing interests.

### Funding

This work was supported by the CUDAN ERA Chair project at Tallinn University, funded by the European Union's Horizon 2020 research and innovation programme (grant agreement No. 810961). M.M.'s research visit, from which this study originated, was supported by the Dora Plus programme, funded by the European Regional Development Fund and the Republic of Estonia.

### Authors' contributions

M.M.: Conceptualization, methodology, data curation, formal analysis, visualization, writing – original draft, writing – review & editing, project administration. P.S.D.: Conceptualization, resources, methodology (early stage), supervision. M.S.: Conceptualization, methodology, formal analysis, visualization, writing – original draft, writing – review & editing, supervision. All authors read and approved the final manuscript.

Supplementary information

| Language | Keyword (1-gram) | Language | Keyword (1-gram) |
|---|---|---|---|
| **Arabic** | أوكرانيا | **Korean** | 우크라이나 |
| **Catalan** | Ucraïna | **Norwegian** | Ukraina |
| **Czech** | Ukrajina | **Persian** | اوکراین |
| **Danish** | Ukraine | **Polish** | Ukraina |
| **Dutch** | Oekraïne | **Portuguese** | Ucrânia |
| **English** | Ukraine | **Romanian** | Ucraina |
| **Estonian** | Ukraina | **Russian** | Украина |
| **Finnish** | Ukraina | **Serbian** | Украјина |
| **French** | Ukraine | **Spanish** | Ucrania |
| **German** | Ukraine | **Swedish** | Ukraina |
| **Greek** | Ουκρανία | **Turkish** | Ukrayna |
| **Hungarian** | Ukrajna | **Ukrainian** | Україна |
| **Indonesian** | Ukraina | **Urdu** | یوکرین |
| **Italian** | Ucraina | **Vietnamese** | Ukraina |

**Table S1. Chosen keywords referring to Ukraine in 28 languages.**

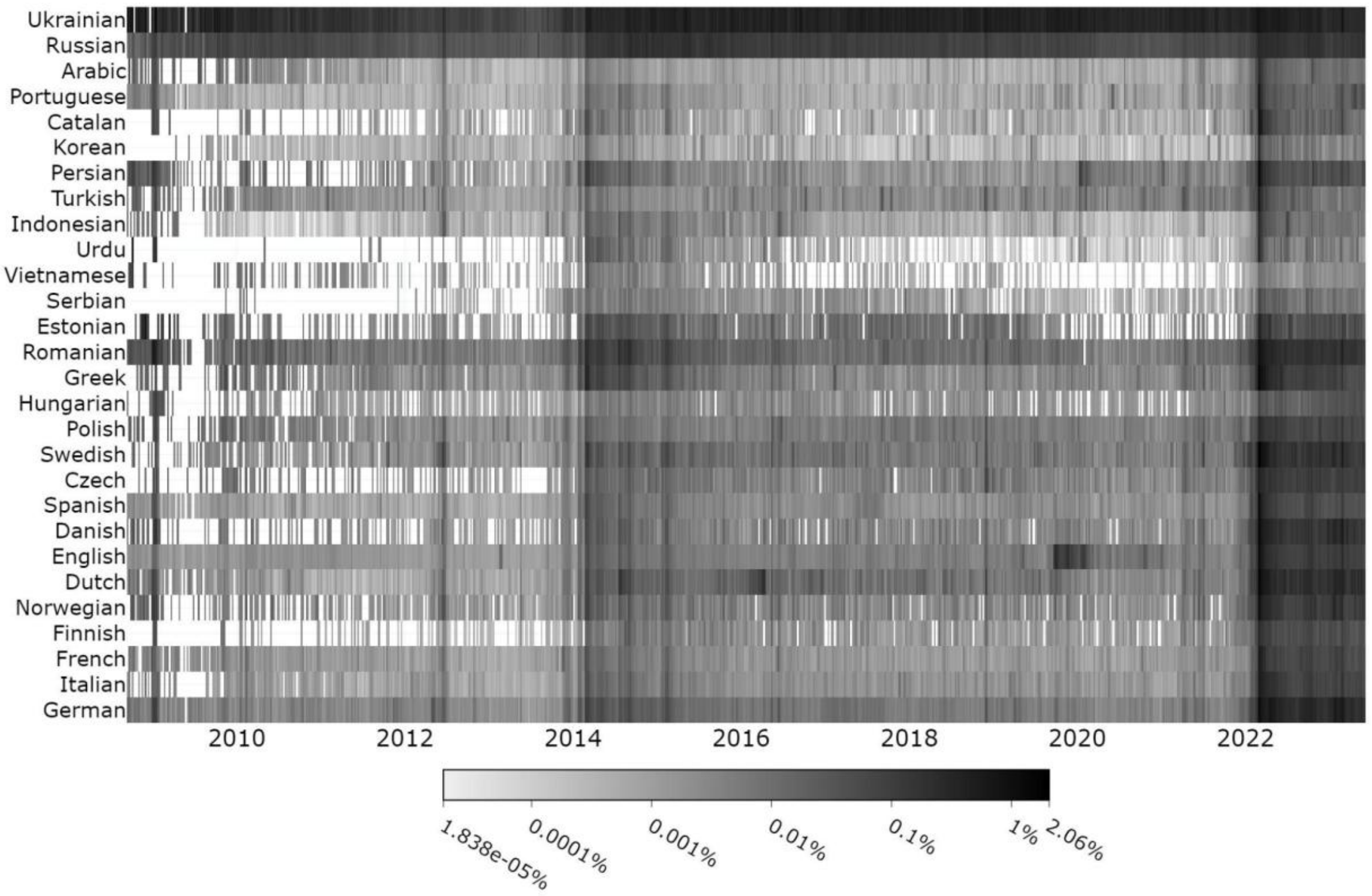

**Fig. S1. Log scale relative frequencies in each language showing a less detailed yet informative overview of attention.** Colorscale corresponds to the height of the bars in Figs. 1 and 2. The colorscale shows the highest relative frequency totalling 2.06% in one week.

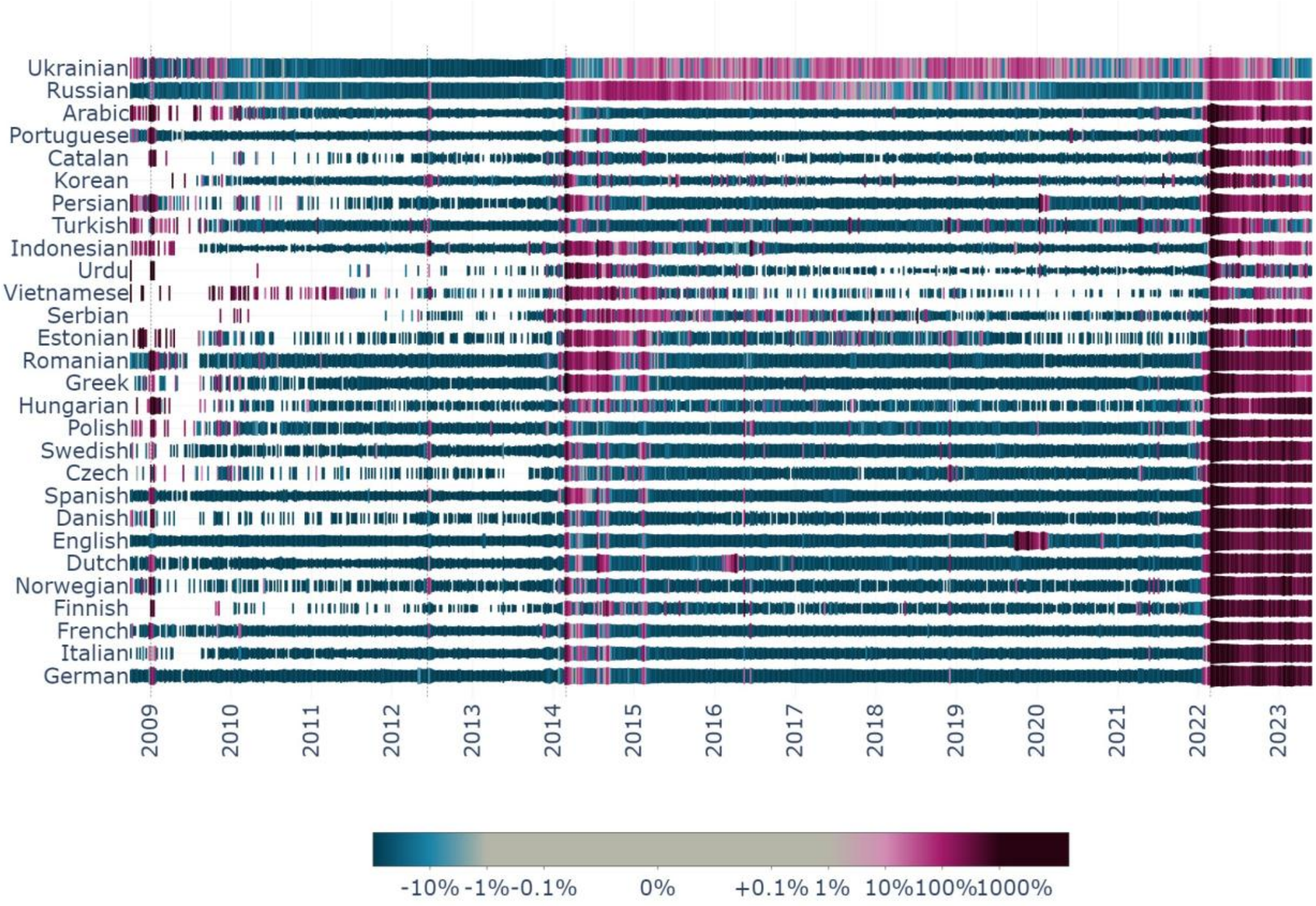

**Fig. S2. Log over/under-expression within language without smoothing.** Corresponds to Fig. 1 without 3 weeks smoothing.

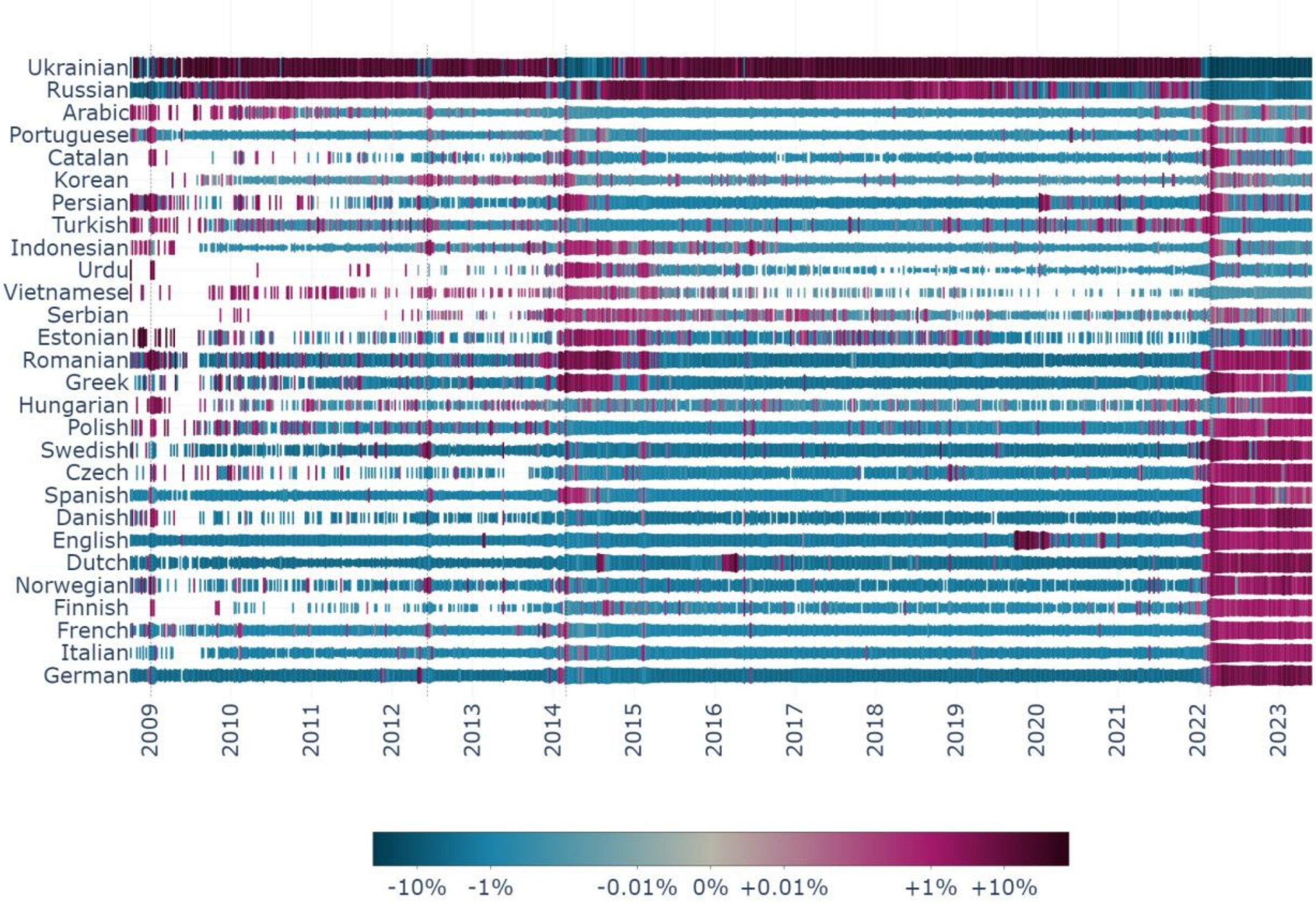

**Fig. S3. Log over/under-expression of "market share" without smoothing.** Corresponds to Fig. 2 without 6 weeks smoothing.

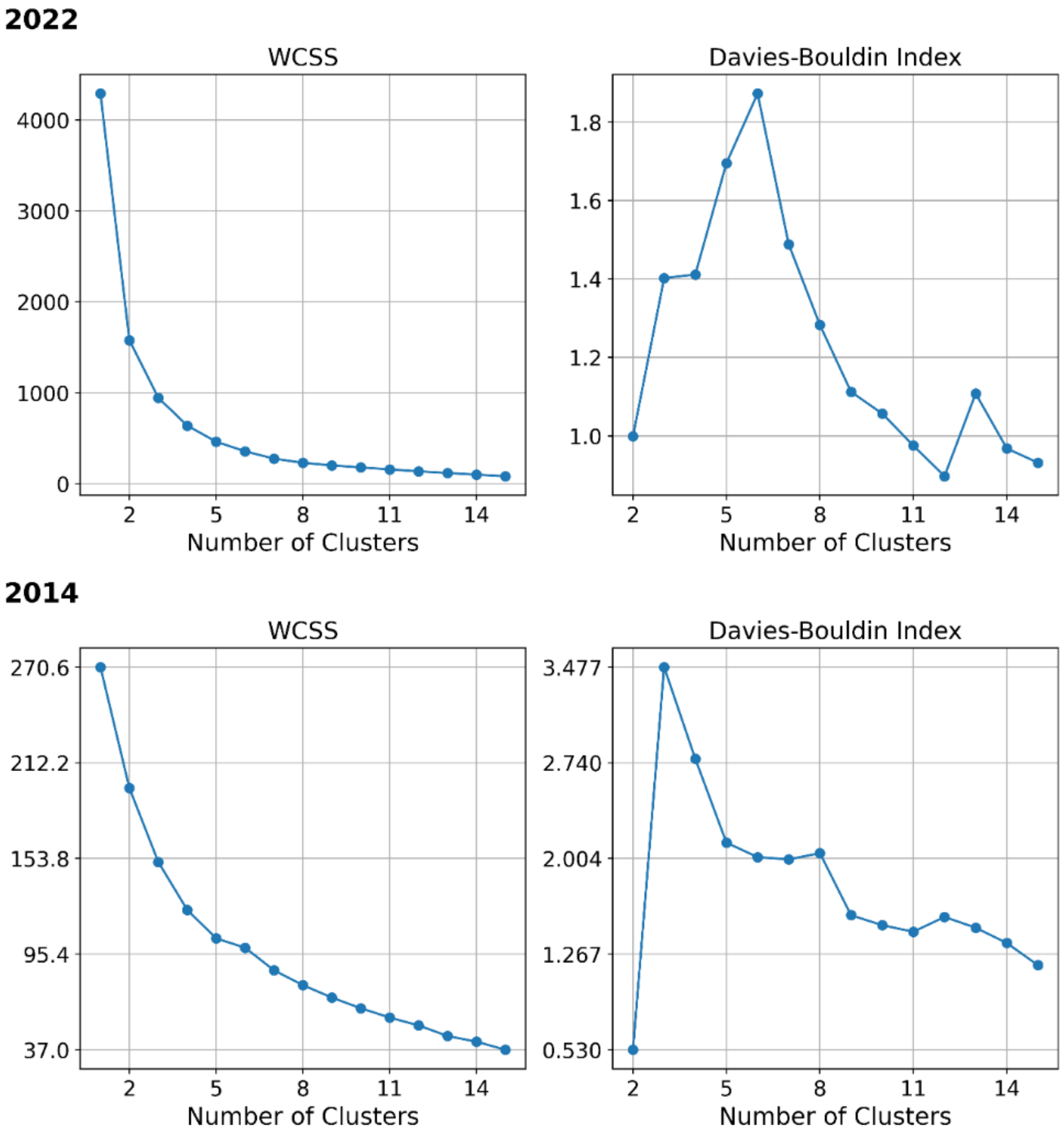

**Fig. S4. Cluster evaluations for 2014 and 2022 invasion time-series in Figure 3.** Left: within-cluster sum of squares (WCSS) for 1–15 clusters; right: Davies–Bouldin index for 2–15 clusters (lower values indicate better separation). Neither metric shows a clear optimum; we chose 5 clusters for 2014 and 6 for 2022 by qualitative assessment of the resulting trajectories.

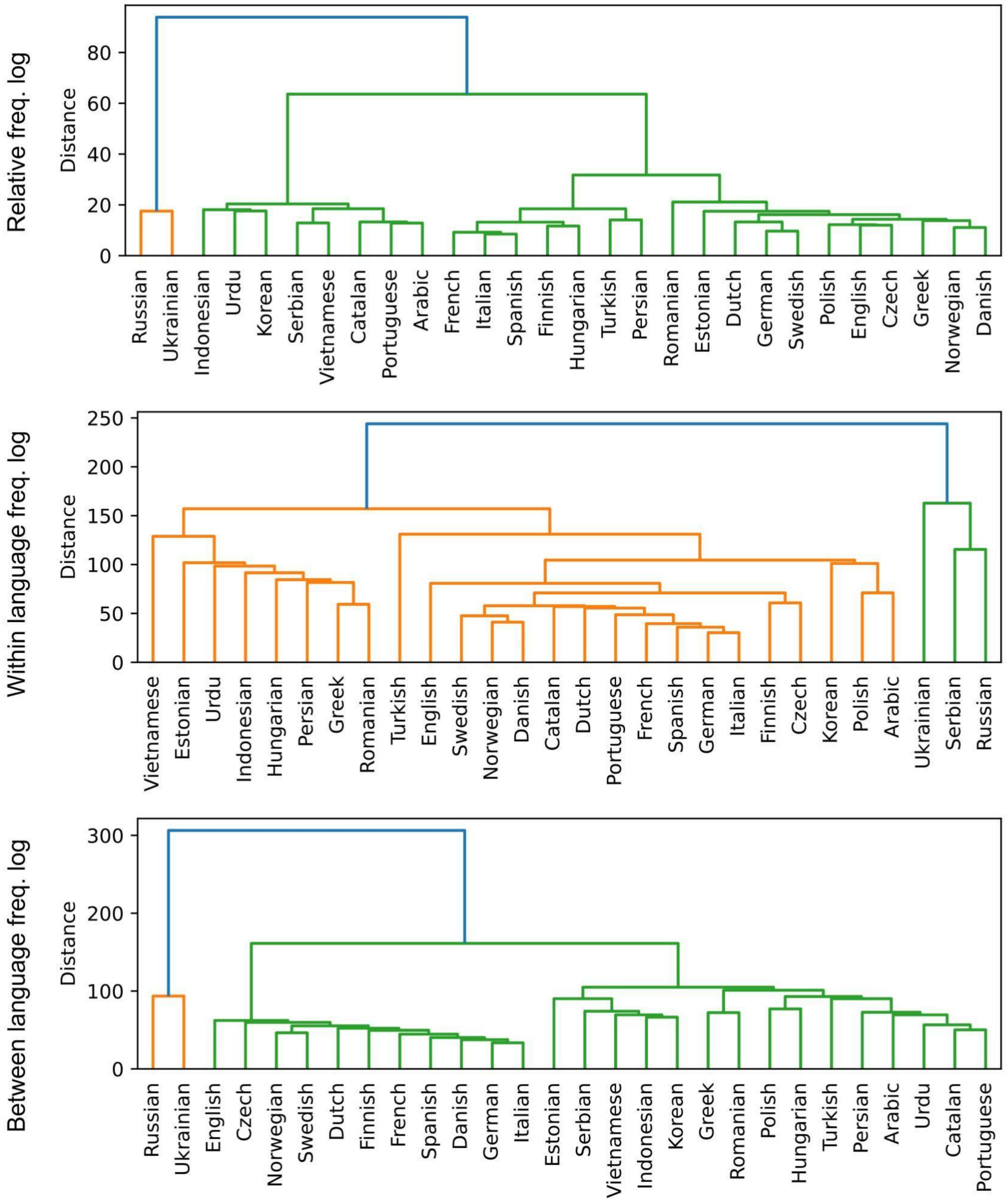

**Fig. S5. Dendrograms demonstrate macro-scale language clustering.** Dendrograms of language similarities across time using Ward metric. Top plot is based on relative frequencies (as seen in Fig. S1). Middle one on over/under-expression within-language as in Figure 1 and the bottom is over/under-expression between-languages based on Figure 2.

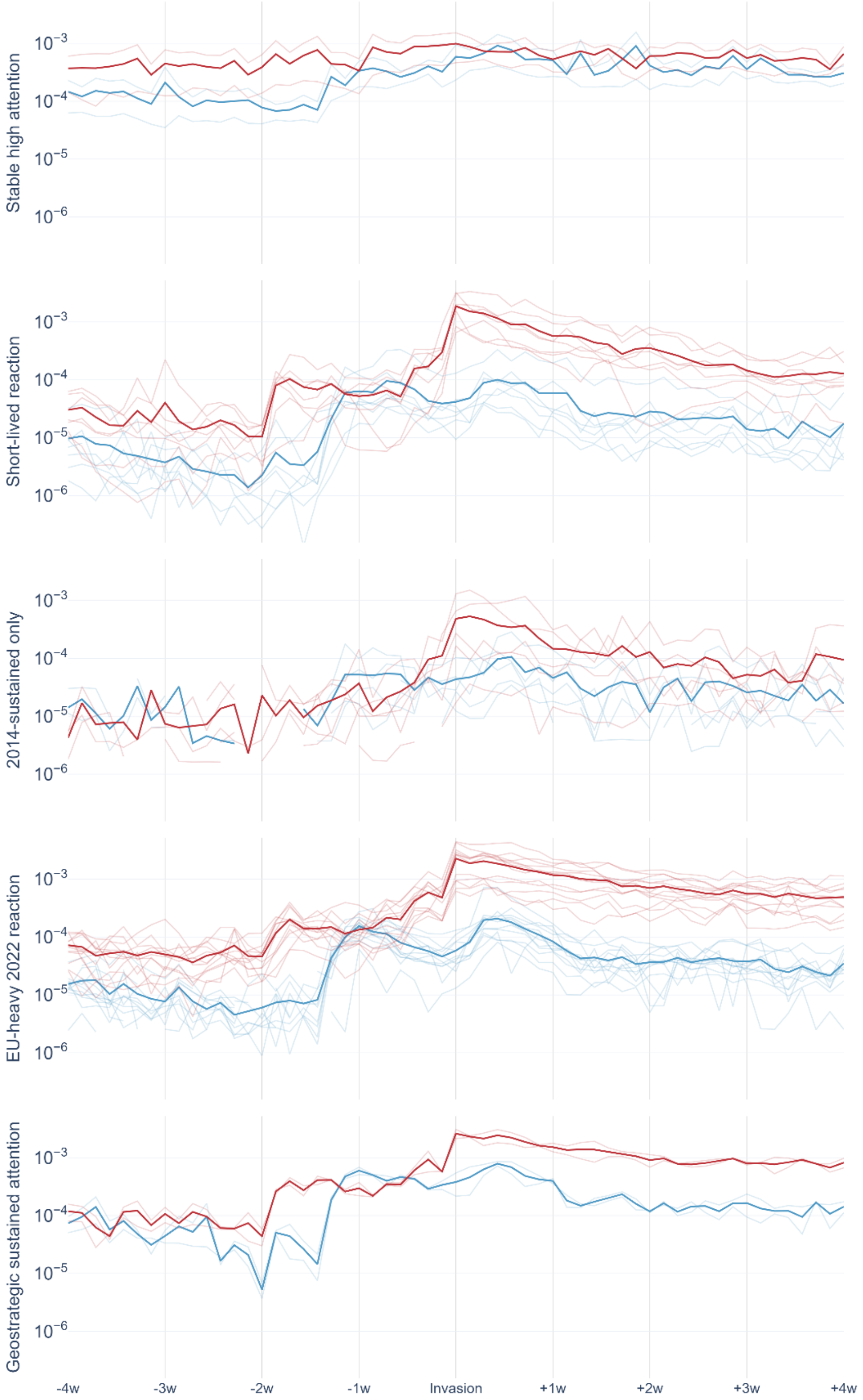

**Fig. S6. Five macro-scale clusters of languages for the 2014 and 2022 comparison, reflecting Figure 4.** Each panel shows the relative frequency of one of five manually assigned clusters; thick lines are the cluster mean for 2014 (blue) and 2022 (red), thin lines individual languages, on a shared logarithmic y-axis. Both years span an 8-week window aligned on their invasion date, marked "Invasion" at the centre, 27 February 2014 (Crimea seizure) and 24 February 2022, as in Figure 4. All 28 languages are shown; Persian and Spanish, discussed in the text as outliers, appear in the Short-lived and EU-heavy panels respectively.

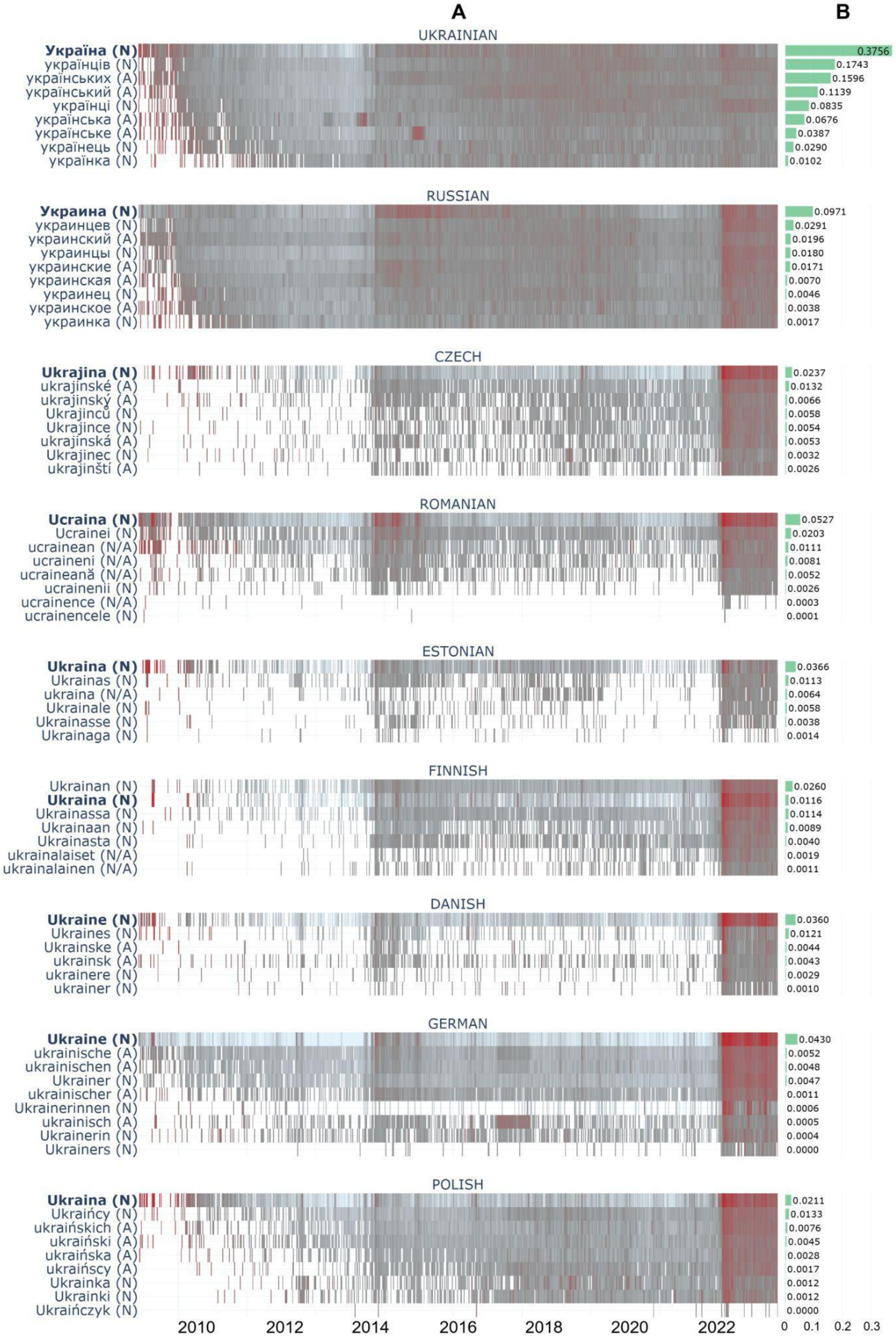
A
B
UKRAINIAN
Україна (N)
українців (N)
українських (A)
український (A)
українці (N)
українська (A)
українське (A)
українець (N)
українка (N)
0.3756
0.1743
0.1596
0.1139
0.0835
0.0676
0.0387
0.0290
0.0102
RUSSIAN
Украина (N)
украинцев (N)
украинский (A)
украинцы (N)
украинские (A)
украинская (A)
украинец (N)
украинское (A)
украинка (N)
0.0971
0.0291
0.0196
0.0180
0.0171
0.0070
0.0046
0.0038
0.0017
CZECH
Ukrajina (N)
ukrajinské (A)
ukrajinský (A)
Ukrajinců (N)
Ukrajince (N)
ukrajinská (A)
Ukrajinec (N)
ukrajinští (A)
0.0237
0.0132
0.0066
0.0058
0.0054
0.0053
0.0032
0.0026
ROMANIAN
Ucraina (N)
Ucrainei (N)
ucrainean (N/A)
ucraineni (N/A)
ucraineană (N/A)
ucrainenii (N)
ucrainence (N/A)
ucrainencele (N)
0.0527
0.0203
0.0111
0.0081
0.0052
0.0026
0.0003
0.0001
ESTONIAN
Ukraina (N)
Ukrainas (N)
ukraina (N/A)
Ukrainale (N)
Ukrainasse (N)
Ukrainaga (N)
0.0366
0.0113
0.0064
0.0058
0.0038
0.0014
FINNISH
Ukrainan (N)
Ukraina (N)
Ukrainassa (N)
Ukrainaan (N)
Ukrainasta (N)
ukrainalaiset (N/A)
ukrainalainen (N/A)
0.0260
0.0116
0.0114
0.0089
0.0040
0.0019
0.0011
DANISH
Ukraine (N)
Ukraines (N)
Ukrainske (A)
ukrainsk (A)
ukrainere (N)
ukrainer (N)
0.0360
0.0121
0.0044
0.0043
0.0029
0.0010
GERMAN
Ukraine (N)
ukrainische (A)
ukrainischen (A)
Ukrainer (N)
ukrainischer (A)
Ukrainerinnen (N)
ukrainisch (A)
Ukrainerin (N)
Ukrainers (N)
0.0430
0.0052
0.0048
0.0047
0.0011
0.0006
0.0005
0.0004
0.0000
POLISH
Ukraina (N)
Ukraińcy (N)
ukraińskich (A)
ukraiński (A)
ukraińska (A)
ukraińscy (A)
Ukrainka (N)
Ukrainki (N)
Ukraińczyk (N)
0.0211
0.0133
0.0076
0.0045
0.0028
0.0017
0.0012
0.0012
0.0000
2010
2012
2014
2016
2018
2020
2022
0
0.1
0.2
0.3

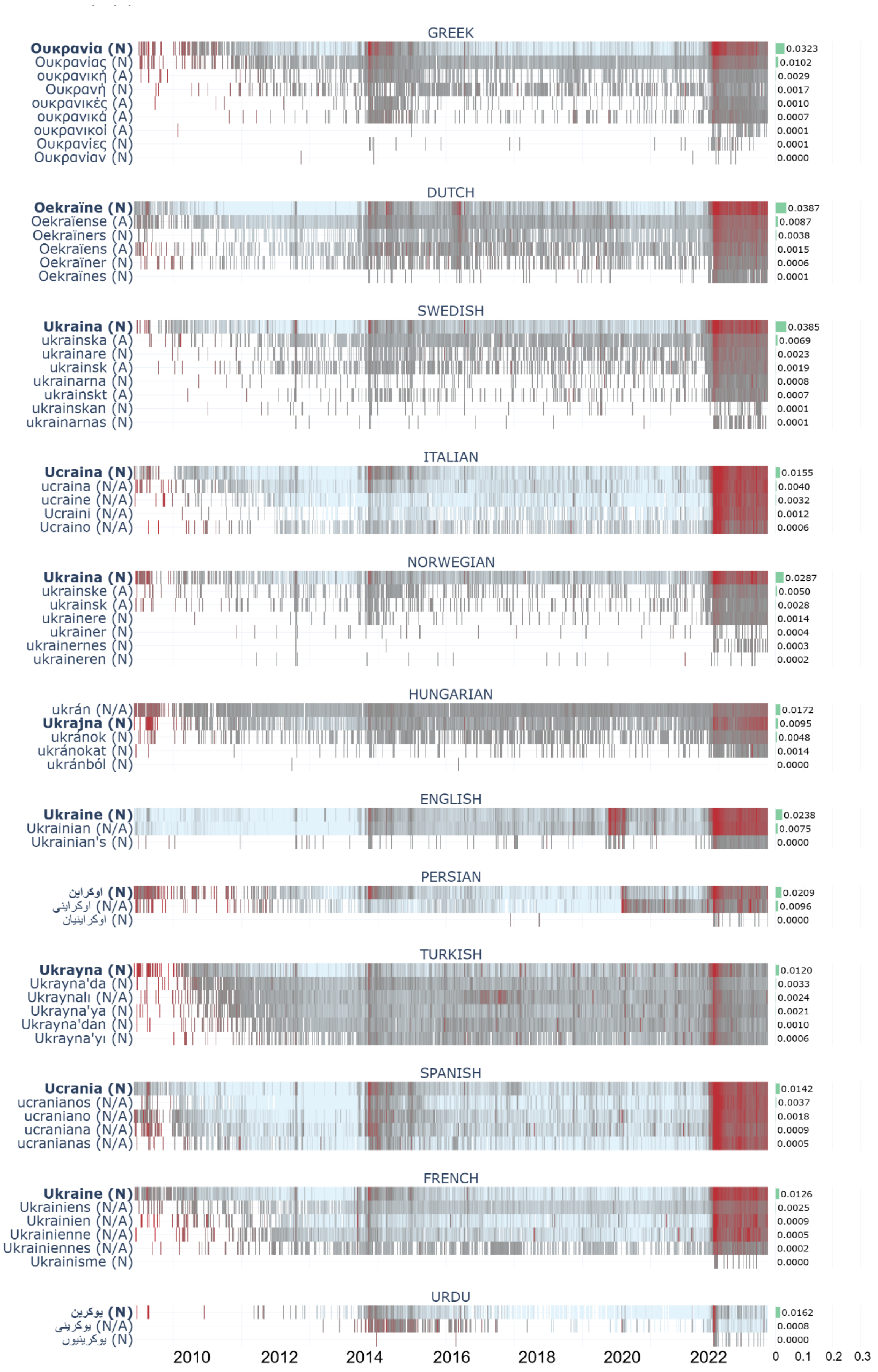
GREEK
Ουκρανία (N)
Ουκρανίας (N)
ουκρανική (A)
Ουκρανή (N)
ουκρανικές (A)
ουκρανικά (A)
ουκρανικοί (A)
Ουκρανίες (N)
Ουκρανίαν (N)
0.0323
0.0102
0.0029
0.0017
0.0010
0.0007
0.0001
0.0001
0.0000
DUTCH
Oekraïne (N)
Oekraïense (A)
Oekraïners (N)
Oekraïens (A)
Oekraïner (N)
Oekraïnes (N)
0.0387
0.0087
0.0038
0.0015
0.0006
0.0001
SWEDISH
Ukraina (N)
ukrainska (A)
ukrainare (N)
ukrainsk (A)
ukrainarna (N)
ukrainskt (A)
ukrainskan (N)
ukrainarnas (N)
0.0385
0.0069
0.0023
0.0019
0.0008
0.0007
0.0001
0.0001
ITALIAN
Ucraina (N)
ucraina (N/A)
ucraine (N/A)
Ucraini (N/A)
Ucraino (N/A)
0.0155
0.0040
0.0032
0.0012
0.0006
NORWEGIAN
Ukraina (N)
ukrainske (A)
ukrainsk (A)
ukrainere (N)
ukrainer (N)
ukrainernes (N)
ukraineren (N)
0.0287
0.0050
0.0028
0.0014
0.0004
0.0003
0.0002
HUNGARIAN
ukrán (N/A)
Ukrajna (N)
ukránok (N)
ukránokat (N)
ukránból (N)
0.0172
0.0095
0.0048
0.0014
0.0000
ENGLISH
Ukraine (N)
Ukrainian (N/A)
Ukrainian's (N)
0.0238
0.0075
0.0000
PERSIAN
اوکراین (N)
اوکراینی (N/A)
اوکراینیان (N)
0.0209
0.0096
0.0000
TURKISH
Ukrayna (N)
Ukrayna'da (N)
Ukraynalı (N/A)
Ukrayna'ya (N)
Ukrayna'dan (N)
Ukrayna'yı (N)
0.0120
0.0033
0.0024
0.0021
0.0010
0.0006
SPANISH
Ucrania (N)
ucranianos (N/A)
ucraniano (N/A)
ucraniana (N/A)
ucranianas (N/A)
0.0142
0.0037
0.0018
0.0009
0.0005
FRENCH
Ukraine (N)
Ukrainiens (N/A)
Ukrainien (N/A)
Ukrainienne (N/A)
Ukrainiennes (N/A)
Ukrainisme (N)
0.0126
0.0025
0.0009
0.0005
0.0002
0.0000
URDU
یوکرین (N)
یوکرینی (N/A)
یوکرینیوں (N)
0.0162
0.0008
0.0000
2010
2012
2014
2016
2018
2020
2022
0
0.1
0.2
0.3

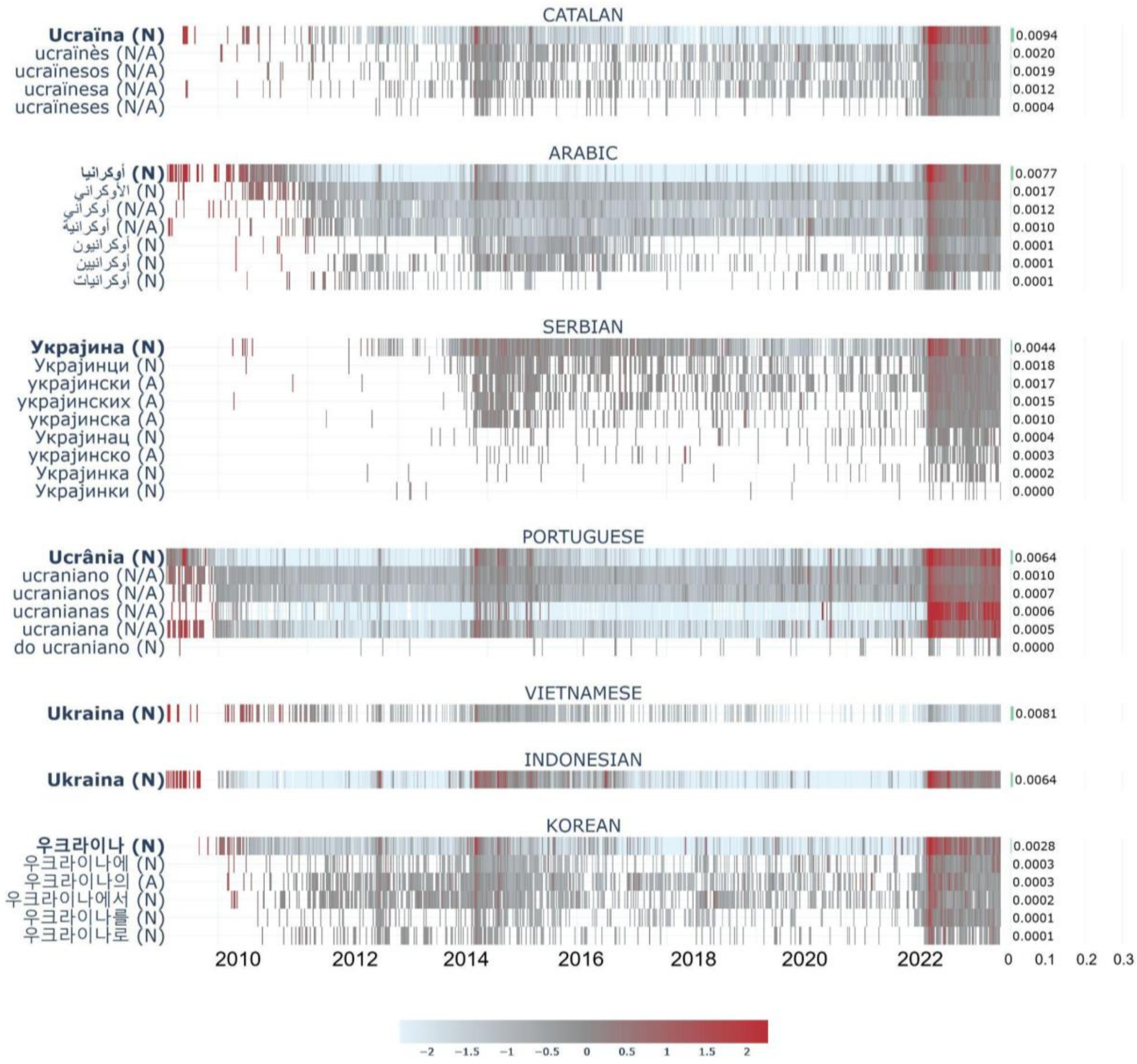

**Fig. S7. Comparison of queried keywords to other word forms.** (A) is adjective, (N) noun and (N/A) both. **A,** log scale word frequencies across time, normalized by average in each language (z-score) with red values showing over- and white values under-expression. At large, the patterns match the chosen noun, with smaller exceptions. **B,** sum of frequencies of keywords in relation to all other words in that language, with shared scale across languages and sorted from most to least frequent. In all languages except Hungarian and Finnish, the chosen form of noun has the highest frequency.